# Strong tuning of Rashba spin orbit interaction in single InAs nanowires

*Dong Liang, Xuan P.A. Gao*[*]

Department of Physics, Case Western Reserve University, Cleveland, Ohio, 44106, USA

[*]Email: xuan.gao@case.edu

**Abstract**

A key concept in the emerging field of spintronics is the gate voltage or electric field control of spin precession via the effective magnetic field generated by the Rashba spin orbit interaction. Here, we demonstrate the generation and tuning of electric field induced Rashba spin orbit interaction in InAs nanowires where a strong electric field is created either by a double gate or a solid electrolyte surrounding gate. In particular, the electrolyte gating enables six-fold tuning of Rashba coefficient and nearly three orders of magnitude tuning of spin relaxation time within only 1 V of gate bias. Such a dramatic tuning of spin orbit interaction in nanowires may have implications in nanowire based spintronic devices.

Nanowires of small band gap III-V semiconductors such as indium arsenide (InAs) have recently attracted significant interest in nanoelectronic device research.[1-10] With high electron mobility, nanostructures of InAs have appealing potential for high speed electronics such as field effect transistor (FET).[3-8] In addition to high electron mobility that is important for charge transport based electronic devices, InAs also has strong spin-orbit interaction (SOI) and was proposed to be an essential material for spintronic devices such as spin FET in which controlling the electrons' spin degree of freedom renders the device's functionality.[11-13] Therefore, InAs nanowire appears to be also an interesting quasi-one dimensional (1D) platform to explore spintronic devices such as 1D spin-FET which exploits the SOI effect to control electrons' spin. SOI is a relativistic effect experienced by electrons (or any charge particles) moving in an electric field *E*. In the rest frame of the electron, *E* is Lorentz transformed into an effective magnetic field which couples to the electron's spin and consequently induces spin precession and relaxation. The electric field in this SOI effect could be from the asymmetry in the intrinsic crystal structure, the asymmetric confinement potential in heterostructure interface, or simply an externally applied gate bias. The SOI in the latter two cases is termed Rashba SOI with an effective Hamiltonian $H_R=\alpha(\sigma\times k)\cdot\hat{E}$ in which $\alpha$, the Rashba SOI coefficient is proportional to $E$ and characterizes the spin-orbit coupling strength, and $\sigma$ and $k$ are the Pauli spin matrices and electron wave vector respectively, $\hat{E}$ is unit vector of electric field. In many spintronic devices such as spin FET, the Rashba SOI is the relevant effect since it renders opportunity to electrically manipulate electron's spin without the need of external magnetic field.

While much initial studies of SOI effects in semiconductors were on planar two-dimensional electron gas (2DEG) systems,[14-18] there is an increasing interest in the SOI effect in quasi-1D quantum wires fabricated from 2DEG, [19, 20] or nanowires synthesized by chemical means.[21-25] Magneto-transport study of the weak anti-localization (WAL) effect was performed in InAs[21-23] and InN nanowires [24] where short spin relaxation times were found, showing the existence of strong SOI. In some experiments,[21, 22] electron spin relaxation length was also found to be tunable by a factor of two via back gate. Besides III-V nanowires, for group IV semiconductor based p-type Ge/Si core/shell nanowires in which the SOI is expected to be small, a surprisingly strong and tunable SOI was observed via the back gate controlled WAL effect. [25] However, the SOI strength decreases with gate voltage, inconsistent with a simple Rashba interpretation of the observed gate tuning effect.[25] Thus a purely electric field induced generation and control of Rashba SOI in nanowire system is still lacking. A drawback in the conventional back or top gated device is that the gate not only imposes a static electric field on the carriers, but also modifies the Fermi momentum (through carrier density change), which both contribute to $H_R$. Furthermore, the small dielectric constant and large thickness of conventional oxide dielectric layer has made it requiring more than 10 V in gate voltage to induce appreciable (e.g. a factor of two in Ref. 21, 22) change in SOI strength. It would be desirable if efficient electric field tuning of Rashba SOI can be achieved in nanowires to enable future spintronic devices with superior performance.

In this work, we have explored two approaches of electric field tuning of Rashba SOI in InAs nanowires. We first use a doubly gate device to demonstrate the purely electric field controlled Rashba SOI in InAs nanowire where Rashba coefficient $\alpha$ is tuned by a factor of two with 30 V between the gates. In contrast to prior experiments involving single gate, our doubly gated devices allow us to create an electric field across the nanowire without changing the electron density. This has allowed us to observe enhanced $\alpha$ for both positive and negative electric fields across the nanowire, attesting to the Rashba origin of SOI. In the second experiment, we demonstrate much more efficient tuning of $\alpha$ in InAs nanowire via gating through a solid electrolyte, polyethylene oxide (PEO) mixed with lithium perchlorate (LiClO$_4$). In the electrolyte gating, a strong electric field ($E\sim10^7$ V/cm) is generated at the surface of nanowire, enabling the six-fold tuning of $\alpha$ (0.5-3×10$^{-11}$eVm) and nearly three orders of magnitude tuning of spin orbit relaxation time $\tau_{so}$ (5×10$^{-14}$s - 2×10$^{-11}$s) at 2 K with only 1 V of gate voltage.

InAs nanowires were grown on silicon substrate capped with 100 nm electron-beam evaporated $Al_2O_3$. Gold nanoparticles with 20 nm or 40 nm diameters (Sigma Aldrich) were applied to the substrate as catalysts. The substrate was placed 13.5 cm away from the center of the furnace. The growth was carried out in 1-inch horizontal tube furnace (Lindberg/Blue M) at $T$=620ºC under the pressure of 1.0 Torr for 1 hour, followed by natural cool down. The nanowires are zinc blende structures along [111] growth direction.[26] To fabricate devices, the as-grown nanowires were sonicated and suspended in ethanol solution which was then dropped onto *p*-type degenerately doped silicon substrate which has 300nm silicon dioxide on surface through which back gate voltage can be applied. Photolithography and electron-beam evaporation were used to define 60 nm thick and 2 μm spaced Ti/Al or Ni electrodes as source and drain contacts. The sample was dipped in 0.5% hydrofluoric acid (HF) for 3 seconds before metal evaporation to remove native oxide and ensure ohmic contacts. To fabricate the top gate in the doubly gated devices, PMMA 950C2 was spin-coated on the chips of nanowires with patterned electrodes, yielding ~1 μm thick PMMA layer on top of the nanowires. Then aluminum top gate electrode was evaporated on top of the PMMA layer to overlay the nanowire channel. To fabricate nanowire FETs with solid electrolyte surrounding gate, the chip with nanowires supported on substrate was dipped in buffer HF for 1 minute after lift-off to partly remove silicon dioxide layer beneath the nanowires so that a small gap was created between the nanowire and substrate to suspend the nanowire. Then $PEO/LiClO_4$(8:1 in weight ratio) dissolved in methanol was spin-coated on the chip at a speed of 1500 rpm and the sample was baked at 90ºC until residual moisture and methanol were removed. Two-terminal nanowire devices were measured by low frequency lock-in technique in a Quantum Design PPMS cryostat under typical source drain bias of 100-200 μV.

The schematic of doubly gated InAs nanowire device is shown as Fig.1a. 300 nm thick $SiO_2$ and ~1 μm PMMA serve as the bottom and top gate dielectrics. The device is similar to standard nanowire FET with both back and top gate except that here a floating voltage $V_{(t-b)g}$ is applied between the gates to induce $E$ across the nanowire without changing the electron density $n$. Similar doubly gated device geometry was used recently to study $E$ induced broken symmetry state in bilayer graphene.[27] The electron density $n$ in nanowire is estimated to be 7000/μm using the gate transfer measurement under the standard backgating configuration (supplemental material). Although the electron density remains unchanged with the application of $E$ under this doubly gated configuration, we note that the conductance $G$ of nanowire decreases when $V_{(t-b)g}$ is increased (Fig.1b). This indicates that there is stronger scattering at the nanowire-PMMA interface when the electrons are attracted towards the PMMA side of the dielectrics at positive $V_{(t-b)g}$. Magneto-transport experiments were performed to elucidate the electric field manipulation of SOI effect. Magneto-conductance, $\Delta G(B)=G(B)-G(B=0)$, measured at various $V_{(t-b)g}$'s within ±1 Tesla perpendicular magnetic field $B$, are shown in Fig.1c for $T$=2 K. At small $V_{(t-b)g}$'s, a positive magneto-conductance due to weak localization (WL) is observed.[28, 29] At large positive or negative voltages between the two gates, a negative magneto-conductance emerges around $B$=0, indicating the WAL effect [21-23, 25] where the strong SOI gives rise to a spin-orbit relaxation length ($l_{so}$) shorter than the phase coherence length ($l_\varphi$) of electrons. This evolution from WL to WAL as $|V_{(t-b)g}|$ increases clearly shows the increased SOI at larger $E$ in our study.

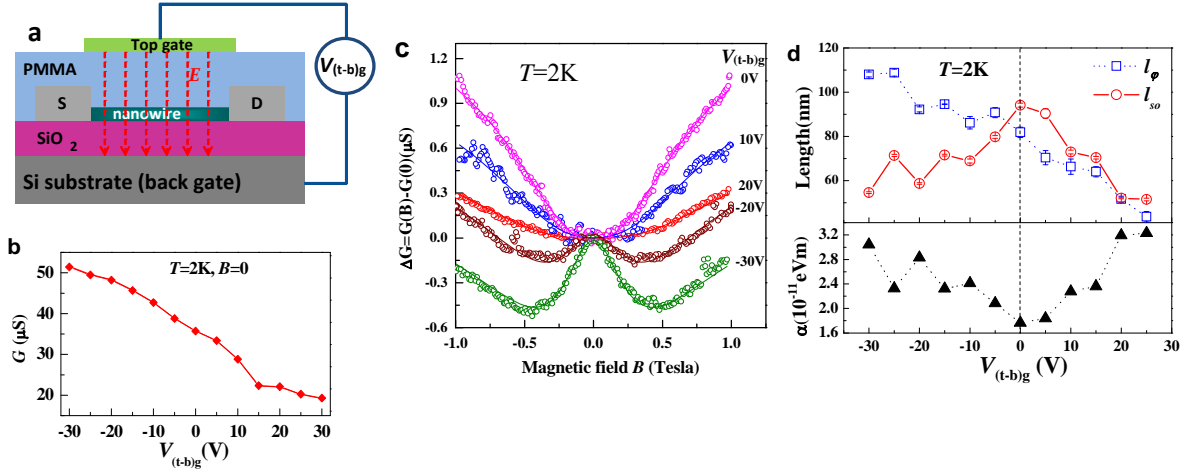

**Figure 1.** (a) A schematic of InAs nanowire device where two (top and bottom) gates are used to apply an electric field across the nanowire for the generation and control of Rashba spin orbit interaction (SOI). (b) Conductance $G$ of an InAs nanowire as a function of $V_{(t-b)g}$, the voltage between the top and bottom gates. (c) Magneto-conductance of an InAs nanowire as a function of perpendicular magnetic field as various $V_{(t-b)g}$'s showing the weak anti-localization effect induced negative magneto-conductance at high $V_{(t-b)g}$ due to the increased Rashba SOI. (d) (top) Spin relaxation length $l_{so}$ and phase coherence length $l_\varphi$ vs. $V_{(t-b)g}$ as obtained from fitting magneto-conductance data to weak anti-localization theory, Eq.(1). (bottom) SOI interaction strength $\alpha$ tuned by the voltage between top and bottom gates.

In the quantum transport of diffusive electrons, WL arises from the constructive quantum interference between time-reversed loops of electron diffusion paths and causes a negative conductance correction.[28,29] In the presence of SOI, electron's spin state is affected by the effective magnetic field induced by SOI thus the negative WL correction to conductance is suppressed. This is the essence of WAL in which the spin-orbit relaxation length $l_{so}$ is another important length scale besides $l_\varphi$. We use the WAL theory of magneto-conductance to quantitatively extract $l_\varphi$ and $l_{so}$ and obtain the SOI strength. In our case, since the mean free path $l\sim18$ nm (supplemental material) is smaller than the nanowire diameter $w=40$ nm, we use the 1D-WAL model in the dirty metal region[21]

$$G(B) = G_0 - \frac{2e^2}{hL}\left[\frac{3}{2}\left(\frac{1}{l_\varphi^2} + \frac{4}{3l_{so}^2} + \frac{1}{l_B^2}\right)^{-1/2} - \frac{1}{2}\left(\frac{1}{l_\varphi^2} + \frac{1}{l_B^2}\right)^{-1/2}\right] \quad (1)$$

Here, $G_0$ is the classical Drude conductance. $L=2$ μm is the length of nanowire. $h$ is the Planck's constant. The magnetic relaxation length is expressed as $l_B = \frac{\sqrt{3}\hbar}{eBw}$. $G_0$, $l_\varphi$ and $l_{so}$ are the fitting parameters in Eq.(1). We restricted the fitting to $B<1$ T to fulfill the small field requirement: $\sqrt{h/eB} > w$. Solid curves in Fig. 1c represent the fitting to the experimental data. The fitted $l_{so}$ and $l_\varphi$ at a number of $V_{(t-b)g}$'s between ±30V are shown in the top panel of Fig. 1d. Note that the 1D-WAL requirement of $l_\varphi > w$ is indeed satisfied as seen from the fitting results. We see that $l_\varphi$ decreases as $V_{(t-b)g}$ increases from negative to positive values. This behavior correlates with the conductance vs. $V_{(t-b)g}$ and is the direct result of the decreased scattering time or diffusion constant at positive $V_{(t-b)g}$. The key result in Fig.1d is that spin-orbit relaxation length $l_{so}$ shows nearly symmetrical decrease around $|V_{(t-b)g}|=0$, independent of the polarity of $V_{(t-b)g}$. Plotting the Rashba SOI coefficient $\alpha = \hbar^2/2m^* l_{so}$ vs. $V_{(t-b)g}$ in the bottom panel of Fig.1d, we see that $\alpha$ has increased by almost two times from $1.6 \times 10^{-11}$ to $3.2 \times 10^{-11} eVm$ from $V_{(t-b)g}=0$ to -30V or +30V. Note that this is the first time such polarization independent increase of $\alpha$ is achieved upon gate voltage increase. Since $\alpha$ is proportional to $|E|$ in the Rashba model, our data provide strong evidence that SOI being tuned by electric field here is indeed the Rashba SOI. In previous studies of WAL in either multiple [21] or single [22, 23] InAs nanowires with only backgating, $l_{so}$ and $\alpha$ did not show such symmetrical response to gate voltage, presumably due to the complications of significantly changed carrier density upon application of $V_{gate}$, which causes drastic changes in multiple factors (Fermi momentum/velocity and the number of 1D sub-bands, etc).

The requirement of tens of volts to tune $\alpha$ in doubly gated device is due to the limited $E$ generated through thick dielectric layers. In the following we demonstrate a much more efficient control of $\alpha$ by electrolyte gating. We use PEO/LiClO$_4$ solid electrolyte as the surrounding gate dielectric in which Li$^+$ or $ClO_4^-$ driven by gate voltage diffuse in a PEO matrix and stop at the nanowire-electrolyte interface. In this gating scheme, all the gate voltage is undertaken over very small distance at the surface of nanowire so that a large $E$ is created (Fig.2a). To ensure that the ionic gate covers all the surface of nanowire, we first suspended InAs nanowire and then applied PEO to create a surrounding gate. In such case, the gate voltage induced $E$ on the nanowire is symmetric along the radial direction. Fig. 2b inset is a SEM image of suspended InAs nanowire obtained by etching the SiO$_2$ substrate underneath nanowire. Due to the large gate capacitance, the PEO/LiClO$_4$ electrolyte gate shows much stronger coupling than conventional back-gate, similar to electrolyte gated carbon nanotube and graphene FETs.[30-32] This is illustrated by the nearly 40 times stronger gate tuning of $G$ by PEO/LiClO$_4$ electrolyte gate compared to that of backgate (Fig. 2b).

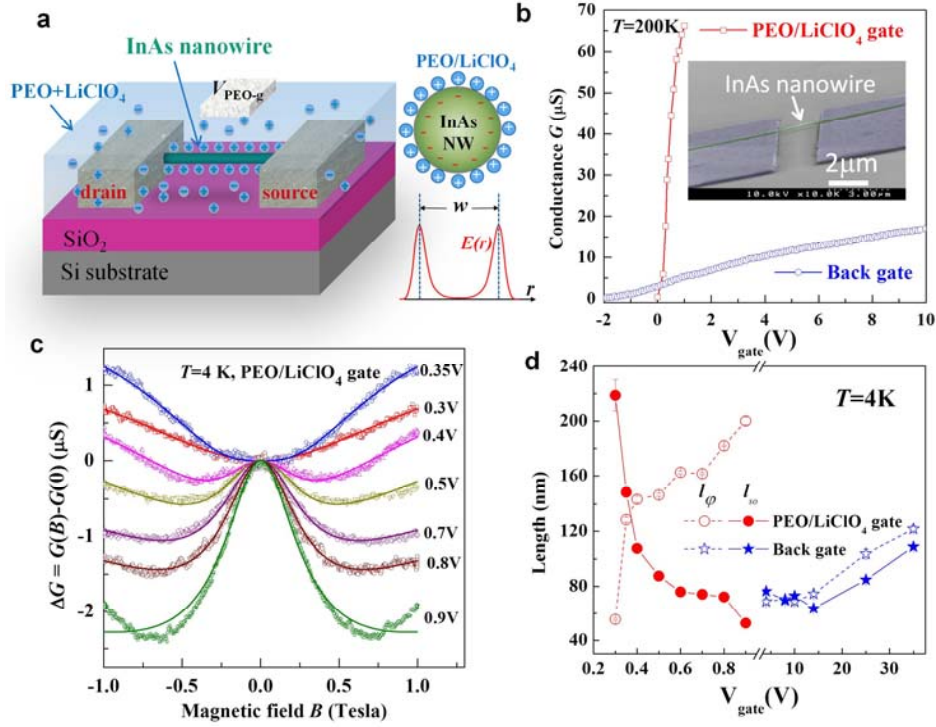

**Figure 2.** (a) Three dimensional schematic view of the suspended InAs nanowire field effect transistor immersed in PEO/LiClO$_4$ solid electrolyte for electrolyte gate control of Rashba SOI. The right panel illustrates the charge and electrical field distribution across the cross-section of nanowire. (b) Conductance versus PEO/LiClO$_4$ surrounding gate and back gate voltages at $T$=200 K, showing the much more effective gate coupling in electrolyte gating. The inset shows a SEM image of suspended InAs nanowire (with PEO removed). (c) The magneto-conductance $G(B)-G(0)$ as a function of perpendicular magnetic field $B$ at $T$=4 K at different PEO/LiClO$_4$ gate voltages. The solid lines are the fitting curves to 1D weak-anti-localization theory. (d) The fitted spin relaxation length $l_{so}$ and phase coherence length $l_{\varphi}$ plotted against gate voltage. The PEO/LiClO$_4$ electrolyte gate is seen to be able to tune spin relaxation length more drastically than the back gate.

Magneto-conductance measurement at low temperature indicates that the electrolyte gating also enables drastic tuning of SOI in InAs nanowire. By increasing PEO gate voltage from 0.3 V to 0.9 V, we observe the magneto-conductance at $T$=4 K transits gradually from positive (WL) to negative (WAL), as shown in Fig.2c, indicating strengthened SOI. $l_{\varphi}$ and $l_{so}$ at $T$=4 K are obtained by fitting data to Eq.(1) and plotted as a function of PEO gate voltage in Fig.2d. What is most striking in Fig.2d is that $l_{so}$ and $l_{\varphi}$ show opposite trend against $V_{\text{gate}}$ although both $l_{so}$ and $l_{\varphi}$ are correlated through the diffusion constant $D$ ($l_{so} = \sqrt{D\tau_{so}}$, $l_{\varphi} = \sqrt{D\tau_{\varphi}}$). The larger $l_{\varphi}$ at higher electrolyte gate voltage is reasonable since positive $V_{\text{gate}}$ adds more electrons into the nanowire and thus enhances the diffusion constant $D$ through the increased Fermi velocity. Observing a reduced $l_{so}$ at the same time would then suggest that increasing PEO gate voltage has drastically shortened the spin-orbit relaxation time, $\tau_{so}$, through enhanced SOI. We find that conventional backgating cannot achieve such strong suppression of $\tau_{so}$ (or enhancement of SOI). For the same device, we measured magneto-conductance at different back-gate voltages. The fitted $l_\varphi$ and $l_{so}$ are included in Fig.2d for a direct comparison of the gate tuning effects between PEO gate and standard backgate. In contrast to PEO gating, backgating induces much weaker change in $l_{so}$ which even shows some increase when the backgate voltage is increased from 15 to 35 V as a result of enhanced $D$. This comparison clearly corroborates the advantage of strong electric field enabled by surrounding electrolyte gate to achieve efficient tuning of SOI.

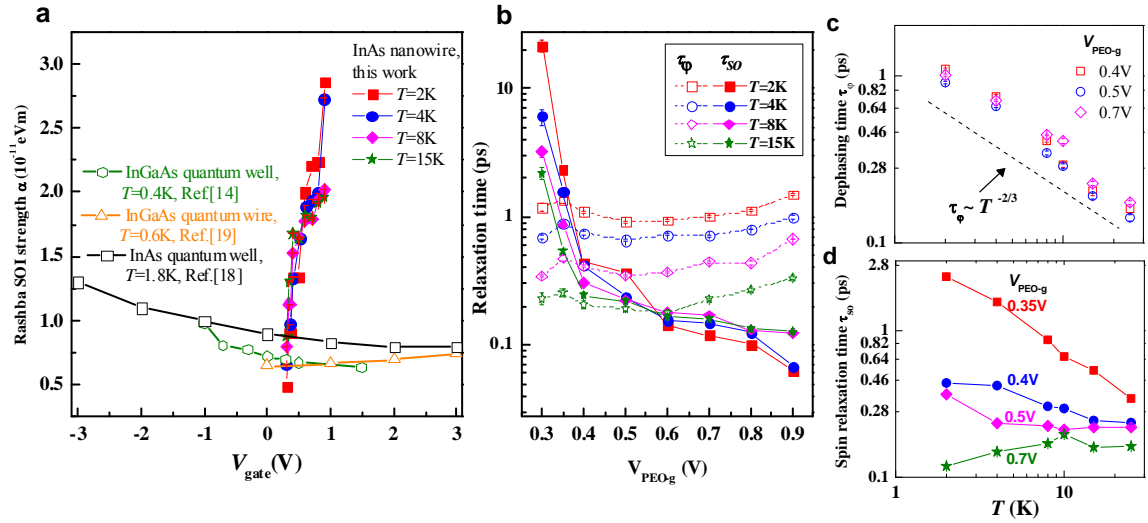

**Figure 3.** (a) Rashba spin orbit interaction strength $\alpha$ as defined by $\alpha = \hbar^2/2m^* l_{so}$ plotted against gate voltage. Our electrolyte gating data are shown as solid symbols for $T$=2, 4, 8, 15 K. Prior experiments on tuning of $\alpha$ by conventional top gate in InGaAs quantum well, [14] InAs quantum well [18] and InGaAs quantum wire [19] are also included for comparison. (b) spin-orbit relaxation time $\tau_{so}$ and dephasing time $\tau_\varphi$ as a function of PEO/LiClO$_4$ gate voltage $V_{PEO-g}$ at $T$=2-15 K. (c, d) $\tau_\varphi$ and $\tau_{so}$ vs. temperature at different $V_{PEO-g}$.

We plot the electrolyte gate voltage tuned Rashba coefficient $\alpha$ vs. gate voltage in Fig. 3a for $T$=2-15 K. Within 1 V of PEO gate voltage, $\alpha$ exhibits six fold increase. In Fig. 3a, we also compare our electrolyte gate tuning of $\alpha$ in InAs nanowire with that obtained via conventional top gating in InGaAs quantum well, [14] InAs quantum well [18] or InGaAs quantum wire [19]. It is clear that solid electrolyte gate enables much more efficient tuning of SOI in InAs nanowire. We point out that the gate tuning of Rashba coefficient in our work is consistent with the known relation between $\alpha$ and the average electric field $\langle E \rangle$: $\alpha = \alpha_0 e \langle E \rangle$. Using $\alpha_0 = 117$ Å$^2$, the Rashba constant for InAs,[33] our data in Fig. 3a would correspond to $\langle E \rangle \sim 10^5$-$10^6$ V/cm, a reasonable value for our case since $E$ is expected to decay rapidly from the peak value (~$10^6$-$10^7$ V/cm) at the nanowire surface toward the center of nanowire due to carrier screening (Fig. 2a).[34] Therefore, we believe that other wrap around gate [3, 35] should work in a similar way to electrolyte surrounding gate as long as the gate is close to nanowire to make *E* large. The strong tunability of SOI in InAs nanowire by PEO gate is further highlighted in the wide dynamic range of spin-relaxation time $\tau_{so}$ varied here. As shown in Fig. 3b, $\tau_{so}$ decreases sharply by nearly three orders of magnitude as $V_{PEO-g}$ increases to 1 V at *T*=2 K. By contrast, the electron dephasing time $\tau_\varphi$ only shows small changes as we increase gate voltage at each temperature. Here, $\tau_{so}$ and $\tau_\varphi$ are calculated from $l_{so}$ and $l_\varphi$ using *D* obtained by the conductance and gate transfer measurements (Supplementary Information). The temperature dependences of $\tau_\varphi$ and $\tau_{so}$ are presented in Fig.3c and Fig.3c for different PEO gate voltages. The dephasing or phase coherent time $\tau_\varphi$ is seen to follow the $T^{-2/3}$ power law, consistent with the dephasing mechanism being the electron-electron collisions with small energy transfers (or the 'Nyquist' dephasing). [28] On the other hand, the temperature dependence of spin relaxation time $\tau_{so}$ weakens as the electrolyte gate voltage or electric field increases. This behavior of $\tau_{so}$ indicates that different spin relaxation mechanisms (e.g. D'yakonov-Perel' or Elliott-Yafet) may be dominating at the small or large Rashba spin-orbit field [12] and is an interesting subject for future study.

In summary, we have employed the double gate and electrolyte surrounding gate methods to induce and tune the Rashba SOI in InAs nanowires. The doubly gated device enables generation of Rashba SOI effect without changing the electron density in nanowire. As such the Rashba coefficient is seen to increase as the electric field strength increases, in both the positive and negative direction. Second, we demonstrate the electrolyte surrounding gate as a highly efficient gating method to tune the Rashba coefficient, due to the strong electric field imposed on nanowire. Our work here shows the promising aspects of gated InAs nanowire devices for nanoscale spintronics.

**Acknowledgements**: D. Liang thanks Keliang He for useful discussion. X. P.A. Gao acknowledges the NSF CAREER Award program (DMR-1151534) and the Donors of the American Chemical Society Petroleum Research Fund (Grant 48800-DNI10) for support of this research.

# Supplementary Information

# Strong tuning of Rashba spin orbit interaction in single InAs nanowires

## Dong Liang, Xuan P.A. Gao[*]

Department of Physics, Case Western Reserve University, Cleveland, Ohio, 44106, USA

[*]Email: xuan.gao@case.edu

1. **Doubly gated InAs nanowire FET characteristics**

We estimate the various characteristics of the InAs nanowire in the doubly gated FET device, as presented in the main text.

The nanowire radius $r$ =20 nm, channel length $L$=2 µm. The thickness of the gate oxide $h$ is 300 nm and the back gate dielectric constant of silicon dioxide is ε=3.9.

Using cylinder on a plane model, the gate capacitance is [1]

$$C_g = 2\pi\varepsilon\varepsilon_0 L/\ln(2h/r)$$

$$C_g = 1.27 \times 10^{-16} F$$

Using conductance $G$ vs. backgate voltage $V_g$ (Fig.S1), the field effect carrier mobility $\mu$ at $V_g$=0 V can be expressed as

$$\mu = \frac{dG}{dV_g}\frac{L^2}{C_g} = \frac{1.5\times10^{-6}}{1}\times\frac{4\times10^{-12}}{1.27\times10^{-16}} \approx 500\, cm^2/V\cdot s$$

We estimate the threshold voltage $V_{th}$=-18 V at $T$=2 K. So line density of the carriers at $V_g$=0 is

$$n = \frac{|V_g - V_{th}|}{e}\frac{C_g}{L} \approx \frac{18}{1.6\times10^{-19}}\frac{1.27\times10^{-16}}{2\times10^{-6}} = 7.14\times10^9 /m$$

The volume density $N$ at $V_g$=0V can be obtained as

$$N = \frac{n}{\pi r^2} = \frac{7.14 \times 10^9}{3.14 \times 4 \times 10^{-16}} = 5.69 \times 10^{24} / m^3$$

If we treat the nanowire as three-dimensional (3D) system (number of occupied one-dimensional (1D) subbands much greater than one), we obtain the 3D Fermi wavenumber

$$k_F = \sqrt[3]{3\pi^2 N} = 5.52 \times 10^8 m^{-1}$$

The corresponding Fermi wavelength is

$$\lambda_F = \frac{2\pi}{k_F} = 1.14 \times 10^{-8} m = 11.4 nm$$

which is about four times smaller than the width (40nm) of nanowire, justifying the 3D assumption.

The Fermi velocity is given by

$$v_F = \frac{\hbar k_F}{m^*} = \frac{1.055 \times 10^{-34} \times 5.52 \times 10^8}{0.023 \times 9.1 \times 10^{-31}} \approx 2.8 \times 10^6 \ m/s$$

The momentum scattering time $\tau$ is calculated as

$$\tau = \frac{\mu m^*}{e} = \frac{0.05 \times 0.023 \times 9.1 \times 10^{-31}}{1.6 \times 10^{-19}} = 6.54 \times 10^{-15} s$$

Mean free path $l = v_F \tau = 18.3 nm$

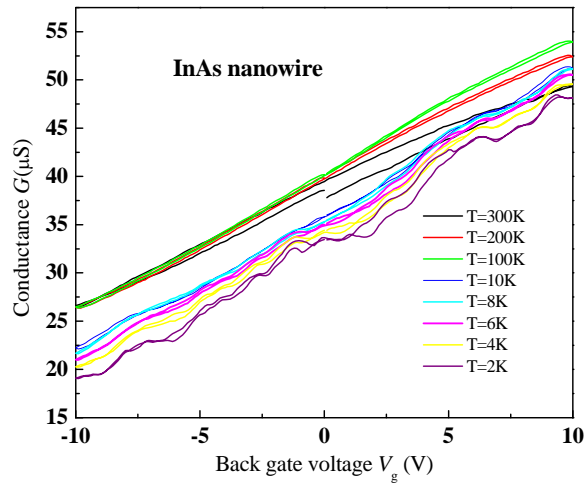

Fig. S1 $G$ vs. backgate voltage $V_g$ at various temperatures for InAs nanowire FET in Fig.1 of main text.

## 2. Electrolyte gated InAs nanowire FET

We estimate the surrounding gate capacitance $C_g$ to be $1.44 \times 10^{-14} F$ by $C_g = 2\pi\varepsilon\varepsilon_0 L/\ln(1+\lambda_D/r)$, where $\varepsilon_0$ is the permittivity, $\varepsilon$ is the dielectric constant of PEO to be ~10, $L$=2 μm and $r$=12.5 nm are the length and radius of nanowire, respectively. We assume Debye length $\lambda_D$ is 1nm. From the trans-conductance $g_m \equiv dG/dV_{PEO}$, one is able to estimate the carrier density $n = (V_{PEO} - V_{th})C_g/eL$ to be 9000/μm and mobility $\mu = g_m \times L^2/C_g$ to be about $600\, cm^2/V \cdot s$ at $V_{PEO\text{-}g}$=0.4 V.

With the carrier density and mobility, we calculate the momentum scattering time $\tau$ and Fermi velocity $v_F$ similar to the previous section. Then the dephasing time $\tau_\varphi$ and spin-orbit relaxation time $\tau_{so}$ for electrolyte gated InAs nanowire are calculated as $\tau_\varphi = l_\varphi^2/D$ and $\tau_{so} = l_{so}^2/D$, with the diffusion constant given by $D = v_F^2 \tau/3$.

To illustrate the strong gate tuning of PEO/LiClO$_4$, we make comparison between electrolyte and back gate for the same device. Fig S2 show the experimental magneto-conductance $G(B)$-$G(0)$ at various back gate voltages at $T$=4 K within ±1 Tesla. By increasing back gate voltages from 0 V to 35V, magneto-conductance in perpendicular magnetic field $B$(-1 to 1 Tesla) shows the crossover from weak localization(WL) to weak anti-localization(WAL). However, large back gate voltage up to 35 V is needed to achieve the crossover and still is not able to tune the magneto-conductance to completely negative, which indicates the weaker tuning of SOI than PEO/LiClO$_4$ electrolyte gating shown in Fig. 2c of main text. $l_\varphi$ and $l_{so}$ are fitted by WAL theory and extracted as a function of backgate voltages at $T$=4 K shown by the blue symbols in Fig. 2d of main text.

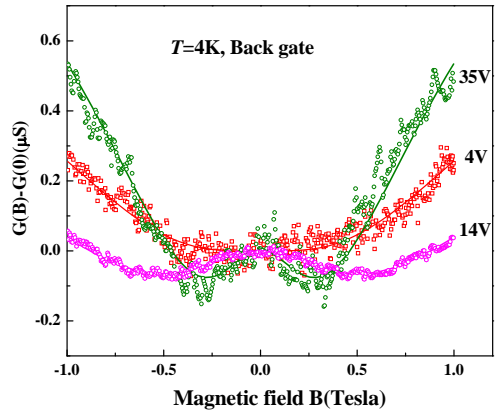

Fig. S2 Magneto-conductance $G(B)$-$G(0)$ as a function of perpendicular magnetic field $B$ at $T$=4 K at different back gate voltages for the device shown in Fig.2 of main text. Solid lines are theoretical fitting by Eq.(1) in main text.